\documentclass[twocolumn,showpacs,amsmath,amssymb,floatfix]{revtex4}

\usepackage{graphicx}% Include figure files
\usepackage{dcolumn}% Align table columns on decimal point
\usepackage{bm}% bold math

\newcommand{\kms}{\ensuremath{\mathrm{km}\,\mathrm{s}^{-1}}}

\newcommand{\LCDM}{$\Lambda$CDM}

\newcommand{\mass}{\ensuremath{{\cal M}}}
\newcommand{\Mst}{\ensuremath{{\cal M}_{\star}}}
\newcommand{\Mg}{\ensuremath{{\cal M}_g}}
\newcommand{\Md}{\ensuremath{{\cal M}_b}}
\newcommand{\Rd}{\ensuremath{R_d}}
\newcommand{\Rp}{\ensuremath{R_p}}

\newcommand{\Sd}{\ensuremath{\Sigma_b}}

\newcommand{\Vb}{\ensuremath{V_{b}}}
\newcommand{\Vh}{\ensuremath{V_h}}
\newcommand{\Vf}{\ensuremath{V_{f}}}
\newcommand{\Vp}{\ensuremath{V_{p}}}

\begin{document}

%\preprint{XXX}

\title{The Balance of Dark and Luminous Mass in Rotating Galaxies}

\author{Stacy S.~McGaugh} 

\affiliation{Department of Astronomy, University of Maryland, 
College Park, MD 20742-2421}    

\date{\today}

\begin{abstract}
A fine balance between dark and baryonic mass is observed in spiral galaxies.
As the contribution of the baryons to the total rotation velocity increases,
the contribution of the dark matter decreases by a compensating amount.
This poses a fine-tuning problem for \LCDM\ galaxy formation models, and may
point to new physics for dark matter particles or even a modification of gravity.
\end{abstract}


%\keywords{dark matter --- galaxies: kinematics and dynamics --- galaxies: spiral} 

\maketitle

%\section{Introduction}

The rotation curves of spiral galaxies become approximately flat at the
largest radii observed \cite{vera,bosma}.  This is one of the strongest
indications of the need for dynamically dominant dark matter in the universe.
While the need for dark matter at large radii is clear, its quantity and
distribution is less so.  

Cogent but contradictory arguments can be made about the relative
importance of dark and luminous mass at small radii in spirals.
The inner shape of rotation curves is well predicted by the distribution
of observed baryons \cite{vAS86,PW}, implying that galactic 
disks are maximal in the sense that they contribute the bulk of the mass
at small radii.  On the other hand, measurements of the velocity dispersions 
of stars perpendicular to galactic disks \cite{bott93,kregel} often suggest 
sub-maximal disks.

Rotation curve shapes measured for spiral galaxies are found to         
correlate strongly with the observed luminosity 
distribution \cite{RFTB,URC,sancisiML,ARMOND}. 
There appears to be a characteristic acceleration 
scale \cite{milgrom83,MDAcc} at which the baryons 
alone can no longer account for the observed dynamics.
The amplitude of the mass discrepancy correlates with acceleration,
providing accurate, albeit empirically motivated, 
determinations of the baryonic masses \cite{MDAcc} of spiral galaxies.
   
This is significant progress in disentangling the dark and baryonic mass.
However, the physics behind these empirical relations is unclear.  
Here I examine the balance between dark and luminous mass in spirals,
and outline the possible interpretations.  

%\section{Correlations Among Galaxy Structural Parameters}

\begin{figure}
\includegraphics[width=0.45\textwidth]{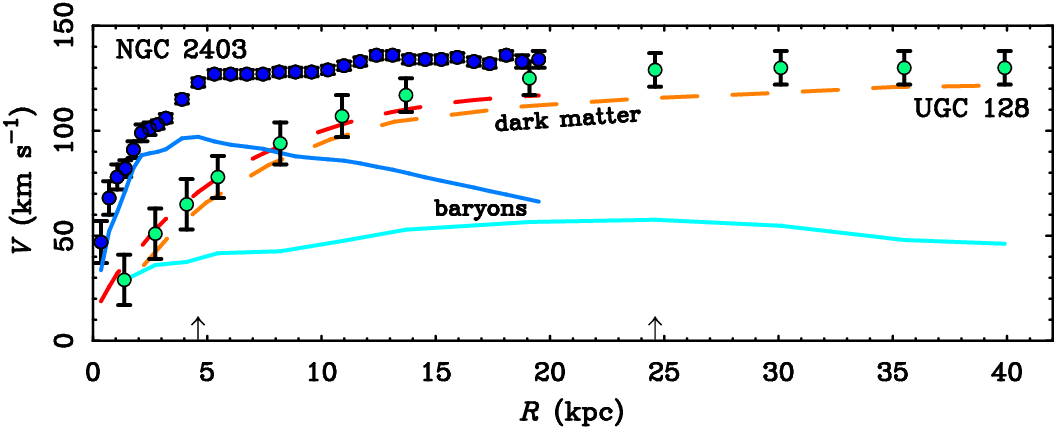}
\caption{\label{fig:N2403U128} (Color online) 
The rotation curves of two galaxies, NGC 2403 and UGC 128, of similar mass
but very different size.  
The contributions of the baryonic components (stars plus gas: solid lines) 
are very different even though the dark matter halos (dashed lines) are similar.  
Lines for the baryons and dark halo of NGC 2403 end with its velocity data 
at 20 kpc while those of UGC 128 continue to 40 kpc.
Arrows mark the radius \Rp\ where the peak of the baryonic rotation occurs.}
\end{figure}

The data used here are tabulated in Ref.~\onlinecite{BTFgood}.  
This is a sample of 
sixty galaxies for which we have extensive, detailed knowledge.  
In particular, their baryonic masses, the sum of stars and gas
\begin{equation}
\Md = \Mst + \Mg,	\label{eqn:mass}
\end{equation}
are well determined.  Examples of the rotation curves of two galaxies 
are shown in Fig.~\ref{fig:N2403U128}.  

Galaxy structural parameters describing the
distribution of baryons are essential to the analysis here.  The azimuthally
averaged light distribution of spiral galaxies can often be approximated as
an exponential disk
\begin{equation}
\Sigma(R) = \Sigma_0 e^{-R/\Rd},	\label{eqn:expdisk}
\end{equation}
where $\Sigma_0$ is the central surface brightness and \Rd\ is the scale length
of the stellar disk.  Such a mass distribution has a rotation curve \cite{F70} 
that peaks at $2.2\Rd$.  In a system with both stars and dark matter, an interesting 
measure is the velocity $V_{2.2}$ at this point \cite{CR,Dutton,Piz} where the 
mass of the stars is most relevant.

Gravity makes no distinction between stars and gas, so a preferable
measure is \Vp, the velocity at the radius \Rp\ where the sum of these baryonic
components make their maximum contribution to the total rotation.
For star dominated galaxies, there is little difference between these quantities.
For gas rich galaxies, which are well represented here, \Rp\ can be considerably
larger than $2.2\Rd$.  \Vp\ usually differs little from $V_{2.2}$ because of the near 
flatness of rotation curves, but in some cases there are significant if modest
differences.

To include the gas as well as the stars, I use the masses tabulated in 
Ref.~\onlinecite{BTFgood} and measure \Rp\ and \Vp\ for each galaxy.
\Rp\ is taken from the combined surface density map (Fig.~\ref{fig:N2403U128}); 
no assumption is made that the disk is purely exponential.  
It is useful to define a baryonic surface mass density
\begin{equation}
\Sd = \frac{3 \Md}{4 \Rp^2}. \label{eqn:surfdens}
\end{equation}
This is equivalent to $\Sigma_0$ for a purely exponential disk with $\Rp = 2.2\Rd$.

\begin{figure}
\includegraphics[width=0.45\textwidth]{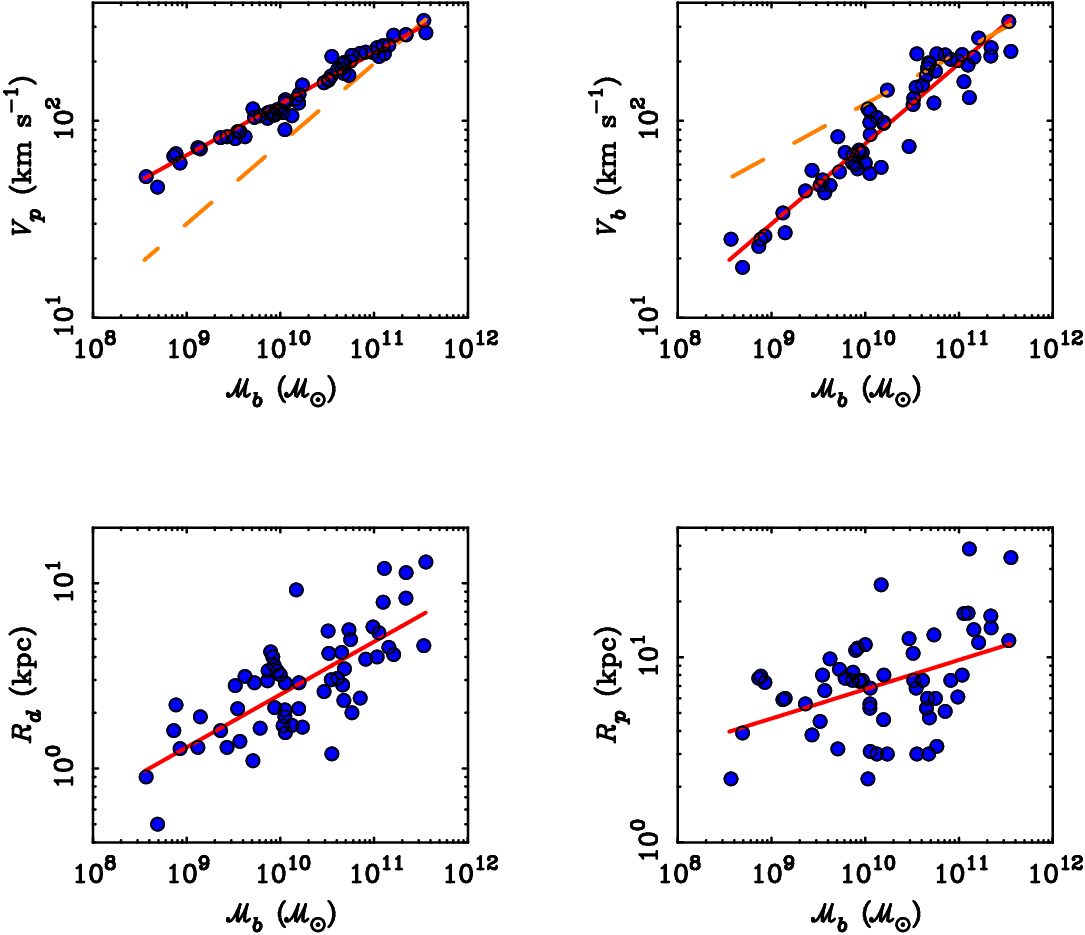}
\caption{\label{fig:VpVbRpMd} (Color online)
Correlations of various galaxy properties with baryonic mass.
Mass is measured in solar masses, velocities in \kms,
and radii in kpc.
Top left: the observed velocity \Vp\ at \Rp.  
Top right: the velocity \Vb\ due
to the baryons at \Rp.  Bottom left: the scale length of the stellar
disk \Rd.  Bottom right: the radius \Rp\ where the baryons
contribute maximally to the rotation.
The solid lines are the fits to the data given in 
Table~\ref{tab:table1}.  For comparison, the fits to
the other velocity is shown as a dashed line in the top panels.}
\end{figure}

The size and rotation velocity of disks correlates with their baryonic mass 
(Fig.~\ref{fig:VpVbRpMd}).
The mass---velocity relation is sometimes referred to as the Baryonic
Tully-Fisher relation \cite{BTForig}.  A number of variations 
\cite{Verh,BdJ} making use of different mass and velocity estimators exist.
The critical difference here from Ref.~\onlinecite{BTFgood} is the consideration
of \Vp\ as the velocity measure as well as the asymptotic flat velocity \Vf.
It is of interest to find how much the baryons contribute to \Vp, so \Vb(\Rp)
is also illustrated in (Fig.~\ref{fig:VpVbRpMd}) for the adopted baryonic mass.
Two size estimators are also shown: the scale length \Rd\ of the stars
and the radius of the peak of the baryonic rotation \Rp.  \Rd\ is the quantity 
traditionally used \cite{CR,Dutton,Piz} when information about the gas is 
not available, and is shown for comparison to the size \Rp\ used here.  
Fits to the data of the form $\log y = a+b \log\Md$ are given 
in Table~\ref{tab:table1}.  

\begin{table}[b]
\caption{\label{tab:table1} Correlations with Baryonic Mass}
\begin{ruledtabular}
\begin{tabular}{ccc}
$y$\footnote{Relations of the form $\log_{10} y = a+b \log_{10} \Md$.}
 %with units specified in Fig.~\ref{fig:VpVbRpMd}.}
 & a & b  \\
\hline
\Vp & $-0.52 \pm 0.03$ & $0.26 \pm 0.01$  \\
\Vf  & $-0.38 \pm 0.05$ & $0.25 \pm 0.01$  \\
\Vb & $-2.19 \pm 0.18$ & $0.41 \pm 0.02$  \\
\Rd & $-2.45 \pm 0.35$ & $0.29 \pm 0.03$  \\
\Rp & $-0.75 \pm 0.44$ & $0.16 \pm 0.04$  \\
\end{tabular}
\end{ruledtabular}
\end{table}

%\section{Residual Correlations}

%\subsection{Velocity and Scale Length Residuals}

A remarkable aspect of the Tully-Fisher \cite{TForig} relation 
(the precursor to the mass---velocity relation) is that it shows no variation with 
scale size \cite{sprayberry,ZwaanTF,hoffman}.  
It is as if the distribution of luminous mass has no impact on the global 
dynamics \cite{MdB98a,Verh}.
Ref.~\onlinecite{CR} proposed using the velocity---scale length residuals
as a measure of the relative importance of baryonic disk and dark matter halo.
The residual from the fitted relations for each object $i$ is 
$\delta y_i \equiv y_i - y(\mass_{d,i})$.  

\begin{figure}
\includegraphics[width=0.45\textwidth]{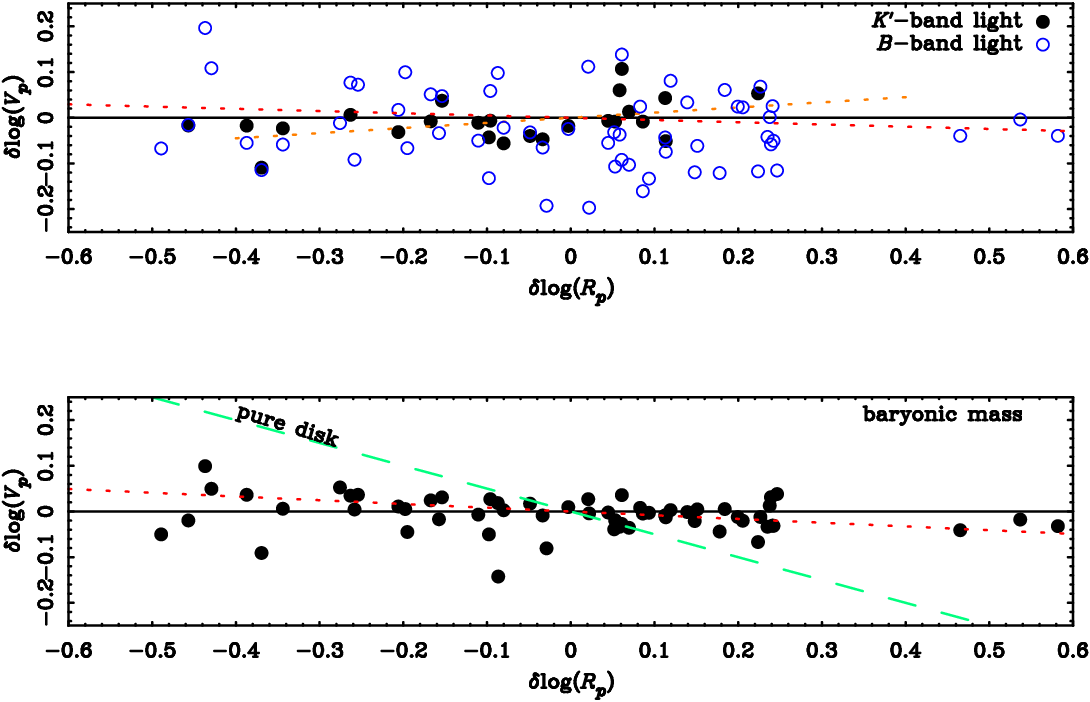}
\caption{(Color online) Velocity residuals from the luminosity-\Vp\ relation (top)
and the \Md-\Vp\ relation (bottom) as a function of scale length residuals.  
In the top panel, open circles are $B$-band data and filled circles are $K'$-band data.  
Fits to these residuals (dotted lines) have slopes close to zero (solid lines).
The mass residuals do not show the slope expected for pure disks (dashed line).
\label{fig:CRresid}}
\end{figure}

The idea is simple:  if the luminous disk contributes significantly to the total mass,
variations in the distribution of baryons should affect the observed velocity.
Imagine taking a compact, high surface brightness galaxy
and stretching it out to become a diffuse, low surface brightness galaxy.
This is essentially what is illustrated in Fig.~\ref{fig:N2403U128}.  
Since $V^2 \propto \mass/R$, the velocity \Vb\ attributable to the baryonic 
component must decline as \Rp\ grows at fixed mass.

\begin{table}
\caption{\label{tab:table2} Residual Slopes}
\begin{ruledtabular}
\begin{tabular}{ccc}
 Estimator & $\partial \log \Vp/\partial \log \Rp$
 	 & $\partial \log \Vf/\partial \log \Rp$ \\
     \hline
$K'$-band  & $+0.11 \pm 0.05$ & $+0.06 \pm 0.04$ \\
$B$-band 	 & $-0.05 \pm 0.05$ & $+0.01 \pm 0.04$ \\
Mass 	 & $-0.08 \pm 0.02$ & $-0.00 \pm 0.01$ \\
\end{tabular}
\end{ruledtabular}
\end{table}

A family of pure disks devoid of dark matter should have $\Vp = \Vb$
and obey a residual relation \cite{CR}
\begin{equation}
\frac{\partial \log \Vp}{\partial \log \Rp} = - \frac{1}{2}.   \label{eqn:resid}
\end{equation}
For a combined system of baryonic disk plus dark matter halo, 
the observed slope will depend on the halo model and
the degree of disk contribution. 
Residuals from both the luminosity---velocity (Tully-Fisher)
and mass---velocity (Baryonic Tully-Fisher) relations are shown
in Fig.~\ref{fig:CRresid}.
Luminosities measured in the $B$-band are available for all galaxies.
$K'$-band photometry \cite{Verh} is available for 24 of the 60 galaxies.  
This near-infrared (2.2 $\mu$m) pass band is thought to
give the closest mapping between stellar light and 
mass \cite{Bell03}, providing a check that the results are not
specific to the particular choice of mass estimator.

No residuals of the type expected for a pure disk are observed.
Indeed, the $B$ and $K'$-bands have slopes of opposite sign,
though neither is significantly different from zero (Table~\ref{tab:table2}).  
The residuals from the mass---flat rotation velocity relation have slope zero.
When \Vp\ is used as the velocity measure instead of \Vf, a slightly negative slope
is inferred.  This is marginally ($4 \sigma$) different from zero.
Given the nature of astronomical observations,
it would be unwise to read too much into the particular value of the slope.
What is clear is that galaxies do not follow the pure disk prediction, having
instead a residual slope very near to zero.

%\section{Discussion}

A tempting interpretation is that all spiral galaxies are dominated by dark matter.  
There is little sensitivity to \Rp\ because the baryonic mass is negligible.
We have estimates of the baryonic mass already, independent of this argument,
so it is interesting to check if this is so.

\begin{figure}
\includegraphics[width=0.45\textwidth]{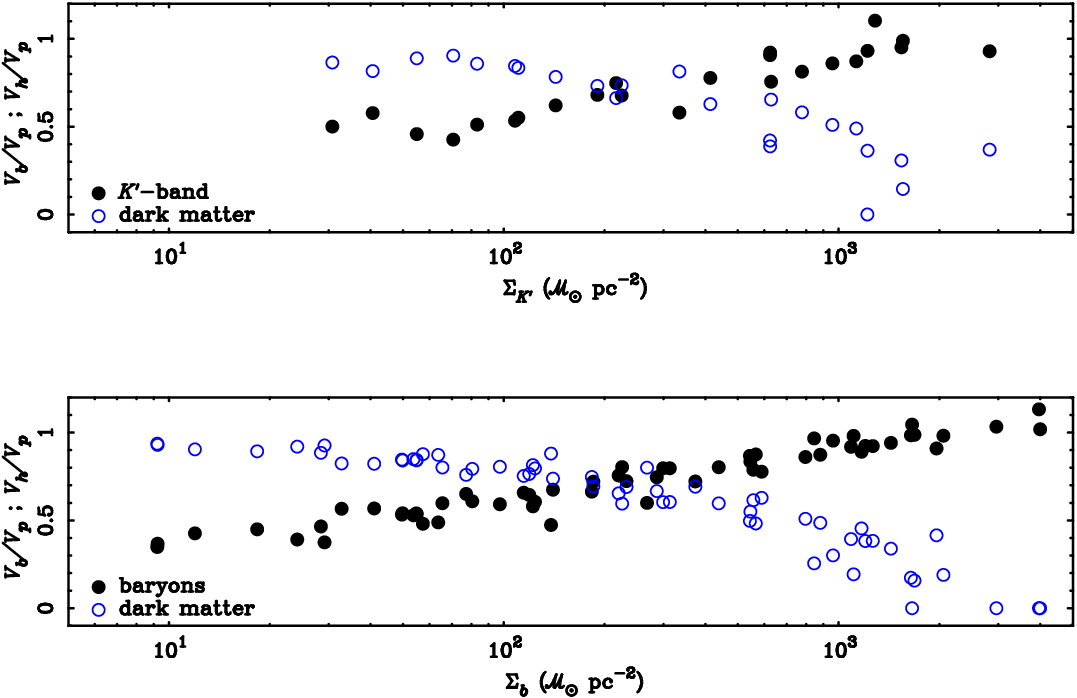}
\caption{(Color online) The fractional contribution to the total velocity at the
peak radius by baryons (\Vb/\Vp) and dark matter (\Vh/\Vp).  These correlate
with baryonic surface density, determined either from the $K'$-band light \cite{Verh}
and stellar population models \cite{Bell03} (top) or equation~(\ref{eqn:surfdens}) 
(bottom).  Disks come closer to being maximal as surface density increases.
Curiously, the dark matter contribution declines in compensation, maintaining a 
see-saw balance with the baryons.
\label{fig:VbVpSd}}
\end{figure}

We can directly compare the contribution of the baryonic (\Vb) and dark halo (\Vh)
components to the total observed velocity \Vp\ at \Rp.
This is shown in Fig.~\ref{fig:VbVpSd} where it can be seen that the fractional
contribution of the baryons, $\Vb/\Vp$, is well correlated with the baryonic
surface density.  This is not specific to the mass estimates adopted here,
being clear also in the $K'$-band data.  For either case, roughly half of the
sample is baryon dominated within \Rp\ ($\Vb > \Vh$).  

Clearly the small slope of the residuals in Fig.~\ref{fig:CRresid} does not 
require sub-maximal disks, as the situation may arise even when some 
disks are maximal.  
The specific relation between the baryon contribution and surface density
(well fit by $\Vb/\Vp = 0.05+0.28 \log \Sd$) depends on the mass estimator,
but such a relation must always be present.  For any choice of
mass estimator, there is a fine-tuning between \Vb\ and \Vh.
$\Vp^2 = \Vb^2+\Vh^2$, yet \Vp\ hardly varies with the distribution of baryons,
while \Vb\ must do so by definition.  This leads to the see-saw balance between
baryons and dark matter apparent in Fig.~\ref{fig:VbVpSd}.  
%This may be related to the long-standing `disk-halo conspiracy' \cite{vAS86}
%wherein \Vf\ remains roughly constant even as $\Vb(R)$ declines.

This balance constitutes a fine-tuning problem for \LCDM.
One does not naively expect the baryonic and dark halo contributions 
to anti-correlate.  In the most basic models \cite{DSS,MMW} one expects
that when the baryons collapse to form the visible galaxy, they draw some of the
dark matter along with them (an effect commonly referred to as adiabatic
contraction \cite{Blumenthal,Oleg,SM05}).  In such a scenario, one would expect
more baryons to mean more dark matter, not less.

The extent to which the observed balance constitutes a fine-tuning problem
is open to debate \cite{MdB98a,vdBD}.  Ref.~\onlinecite{vdBD} suggests that
feedback from the activity of massive stars --- the kinetic energy from stellar 
winds and supernova explosions --- may provide a mechanism to reproduce 
the observations.  It seems strange to invoke such chaotic effects to impart
an organization to the models which is not naturally there. 
Worse, a minority of the baryonic component must strongly affect 
the dominant dark matter --- a case of the tail wagging the dog.  
In order to work, feedback must act in a very specific fashion
which has yet to be realized in sophisticated numerical 
simulations \cite{Fabio,Abadi}.
A completely satisfactory theory of galaxy formation has yet to emerge.

It is worth noting that $\partial \log \Vf/\partial \log \Rp = 0$ 
is an \textit{a priori} prediction of an alternative to dark matter, 
the modified Newtonian dynamics (MOND) \cite{milgrom83}. 
Indeed, while dark matter models struggle with the cusp/core problem
\cite{dBMBR}, MOND fits the rotation curve data in considerable 
detail \cite{ARMOND}.  No fine-tuning is required.

MOND has had other predictions realized,
for example in the dynamics of low surface brightness galaxies \cite{MdB98b}
and in the peak amplitude ratio of the angular power spectrum of the 
cosmic background radiation \cite{boomermond,mondwmap}.
A long standing theoretical objection to MOND has been the lack of a generally
covariant theory, but this obstacle has recently been addressed \cite{TeVeS}.
This in turn may allow the theory to address issues like large scale structure
\cite{skord} about which it has previously been mute.

While MOND has been surprisingly successful, it is not without problems.
The most serious problem facing it at present is the residual mass
discrepancy in clusters of galaxies.  The MOND formula applied to these
systems fails to explain the missing mass problem 
\cite{Aguirre,Sand03,Clowe,PS05} as it does in individual galaxies.  
Perhaps this will prove fatal to the theory, or perhaps there is more
conventional mass in clusters that remains undetected.

There is a third possibility.  We still know very little about the nature of the
dark matter (presuming it exists).  It may possess some property that
imparts the observed balance with baryons in galaxies.
This idea implies a specific interaction between the two that is in some 
way repulsive: the greater the surface density of baryons, the less that
of dark matter.  Such a repulsion would help explain the apparent lack of 
dark matter in high density regions like globular clusters \cite{Scarpa}
and elliptical galaxies \cite{Romanowsky}.  It would also have important 
implications for direct detection experiments.  

No interaction with baryons of the sort envisaged is in the
nature of most hypothesized dark matter candidates.  
Neither cold dark matter nor frequently discussed alternatives like 
warm \cite{dolgovhansen, giudice} or self-interacting \cite{SIDM} 
dark matter do anything of the sort.
Whether it is even possible to endow dark matter with the appropriate 
properties \cite{piazmar} is difficult to say as the possibility has yet to 
be thoroughly explored.  

Irrespective of which type of interpretation may seem preferable ---
the details of galaxy formation, modified dynamics,
or baryon-repulsive dark matter ---
clearly there is important physics at work that has yet to be elucidated.

%\acknowledgements
The author gratefully acknowledges constructive suggestions by the referees.
The work of SSM is supported in part by NSF grant AST0505956.
%AST0206078 and NASA grant NAG513108.

%\bibliography{BTFresidREVTEX}

\end{document}